\documentclass[9pt]{article}
\usepackage{spconf,amsmath,amssymb,graphicx}
\usepackage{bbm}
\usepackage{enumitem}
\usepackage[caption=false,font=footnotesize]{subfig}
\usepackage{hyperref}

\usepackage{xcolor}

\usepackage{glossaries}

\newacronym{rfs}{RFS}{random finite set}
\newacronym{dbscan}{DBSCAN}{density-based spatial clustering of applications with noise}
\newacronym{mcmc}{MCMC}{Markov chain Monte Carlo}
\newacronym{cbmember}{CBMeMBer}{cardinality balanced multi-target multi-Bernoulli}
\newacronym{digsncp}{DI-GSNCP}{doubly inhomogeneous-generalized shot noise Cox process}
\newacronym{ospa}{OSPA}{optimal sub-pattern assignment}

\def\R{\mathbb{R}}

\def\N{\mathbb{N}}

\newcommand{\bs}[1]{\boldsymbol{#1}}

\usepackage{pgfplots,tikz,tikz-3dplot}
\pgfplotsset{compat=1.15}
\usepgfplotslibrary{fillbetween}
\usetikzlibrary{patterns,arrows,plotmarks}
\usepgfplotslibrary{groupplots}
\pgfdeclarelayer{background}
\pgfsetlayers{background,main}
\usetikzlibrary{automata,positioning}
\usetikzlibrary{decorations}
\usetikzlibrary{shapes.arrows}
\usetikzlibrary{shapes.misc}
\usetikzlibrary{tikzmark}
\usetikzlibrary{calc}
\usetikzlibrary{decorations.markings}
\usetikzlibrary{shapes.geometric}
\usepgfplotslibrary{colorbrewer}
\usepgfplotslibrary{colormaps}

\definecolor{color1}{HTML}{0011af}
\definecolor{color2}{HTML}{8819a0}
\definecolor{color3}{HTML}{bf418d}
\definecolor{color4}{HTML}{e37076}
\definecolor{color5}{HTML}{f9a256}
\definecolor{color6}{HTML}{FF0000}
\definecolor{color7}{HTML}{009F6B}
\definecolor{color8}{HTML}{00CC99}

\usepackage{algorithm}
\usepackage{algorithmicx}
\usepackage{algpseudocode}
\makeatletter
\def\BState{\State\hskip-\ALG@thistlm}
\makeatother
\algnewcommand\algorithmicforeach{\textbf{for each}}
\algdef{S}[FOR]{ForEach}[1]{\algorithmicforeach\ #1\ \algorithmicdo}

\makeatletter
\algnewcommand{\LineComment}[1]{\Statex \hskip\ALG@thistlm #1}
\makeatother

\usepackage{cleveref}
\crefname{equation}{Eq.}{Eqs.}
\crefname{figure}{Figure}{Figures.}
\crefname{table}{Table}{Tables}
\crefname{section}{Section}{Sections}

\title{Multi-Sensor Multi-Scan Radar Sensing of Multiple Extended Targets}
\name{Martin V. Vejling$^{\star \dagger *}$, \quad Christophe~A.~N.~Biscio$^{\star}$, \quad Petar Popovski$^{\dagger}$}
\address{$^{\star}$ Dept.~of Mathematical Sciences, Aalborg University, Denmark.\\
$^{\dagger}$ Dept.~of Electronic Systems, Aalborg University, Denmark.\\
* Corresponding author e-mail: martin.vejling@gmail.com}
\begin{document}
\pgfplotsset{
    colormap/winter,
}
%
\maketitle
\begin{abstract}
We propose an efficient solution to the state estimation problem in multi-scan multi-sensor multiple extended target sensing scenarios. We first model the measurement process by a doubly inhomogeneous-generalized shot noise Cox process and then estimate the parameters using a jump Markov chain Monte Carlo sampling technique. The proposed approach scales linearly in the number of measurements and can take spatial properties of the sensors into account, herein, sensor noise covariance, detection probability, and resolution. Numerical experiments using radar measurement data suggest that the algorithm offers improvements in high clutter scenarios with closely spaced targets over state-of-the-art clustering techniques used in existing multiple extended target tracking algorithms.

\end{abstract}
\section{Introduction}
\label{sec:intro}

Extended target sensing refers to estimating the state of a target that with a given sensor can produce multiple detections. This occurs when the extent of the target is large relative to the resolution of the sensor, and it realizes the possibility to not only estimate the position of the target but also its shape. Extended target sensing scenarios are expected to arise with radar sensors relying on future communication systems infrastructure since the resolution is expected to improve due to higher operating frequencies, increased bandwidths, larger array apertures, and deployment of metasurfaces \cite{Bourdoux2020}.

Algorithms for tracking of multiple extended targets with a single sensor using single scan filtering, which refers to sequentially updating estimates through time based on the prediction from the previous time step and newly sampled data, have been developed by adaptation of existing algorithms for point source targets \cite{Granstrom2010:Gaussian,Granstrom2020:ExtendedPMBM}. These methods in general scale exponentially in the number of measurements, hence, clustering is required to group measurements that originate from the same target. This clustering was initially done using methods based on spatial proximity, e.g., distance partitioning \cite{Granstrom2012:Extended}, and density-based spatial clustering of applications with noise (DBSCAN) \cite{Xue2021DBSCAN}. These clustering algorithms cannot properly separate closely spaced targets since the resulting measurements will also be in close proximity, resulting in merging clusters. Attempts to alleviate this have been made by combining clustering algorithms based on different heuristics \cite{Granstrom2012:Extended,Granstrom2012:PHD}, however, the performance is highly dependent on the global clustering hyperparameters which cannot account for inhomogeneity in the sensor properties and is sub-optimal since the model posterior is not considered.

Promising works to overcome the aforementioned challenges have proposed sampling methods for multiple extended target tracking \cite{Granstrom2018:Extended,Bohler2019:Partitioning}, however, these methods do not make explicit target state estimation and instead rely on tracking data association hypotheses, i.e., possible associations between measurements and targets. For this reason, these methods can fail in scenarios where the number of relevant data association hypotheses becomes overwhelmingly large.
In a recent advance \cite{Meyer2021:Tracking}, a sum-product algorithm is proposed for extended target tracking which scales quadratically in the number of measurements. Still, for multi-sensor multi-scan multiple extended target state estimation, this can be too restrictive due to excessively large measurement counts.

When multiple sensors are available, their information should ideally be shared. This has been done decentralized with sub-optimal density fusion \cite{Frohle2020:Decentralized}, but if computational efficiency is not an issue then ideally the information should be combined at a centralized processor \cite{Meyer2018:Scalable,Vo2019:GLMB}. Similarly, information collected across multiple scans should ideally be processed together \cite{Vo2019:Scan}. The combined multi-sensor multi-scan perspective was recently explored in \cite{Moratuwage2022:MultiScan} where the authors propose to leverage the generalized labeled multi-Bernoulli model and addresses multi-dimensional data association hurdles via Gibbs sampling, albeit with the assumption of point source targets.

\begin{figure}[t]
    \centering
    \resizebox{0.47\textwidth}{!}{\input{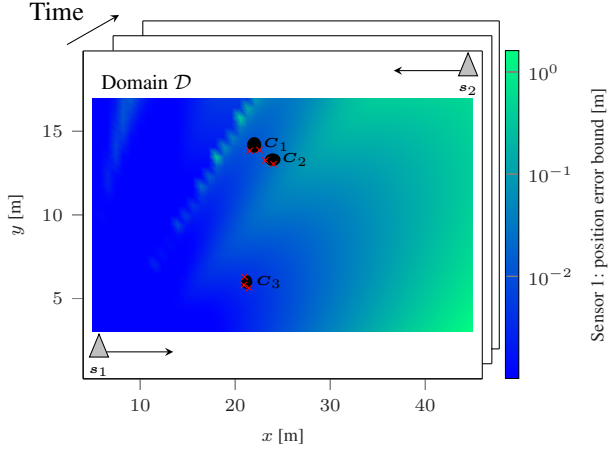}}
    \caption{Illustration of scenario with two sensors and three extended targets. The colormap shows the position error bound with the first sensor.}
    \label{fig:minimal_example}
\end{figure}

The multi-sensor multi-scan multiple extended target sensing scenario is illustrated in \cref{fig:minimal_example} with sensors represented by gray triangles, extended targets by black ellipses, and some of the resolvable detection points by red crosses. Additionally, the position error bound of the first sensor is shown, which summarizes the sensor noise covariance. In this figure, two of the targets are closely spaced while a third target is well-separated, and the clustering algorithms based on spatial proximity will merge the two closely spaced targets into one cluster

To construct an efficient algorithm for target state estimation capable of taking the spatial properties of the sensors into account, we propose a doubly inhomogeneous-generalized shot noise Cox process (DI-GSNCP) to model the multi-sensor multi-scan multiple extended target measurement process. With this model, target state estimation can be done directly by estimating the model parameters, making explicit data association unnecessary. A Bayesian approach is taken, making prior assumptions on the extent and interaction between targets, and we propose an estimation procedure based on a jump \gls{mcmc} sampling method, which scales only linearly in the number of measurements.
The efficacy of the method in high clutter scenarios with closely spaced targets is demonstrated on a radar measurement model.\footnote{Software is provided at \href{https://github.com/Martin497/DI-GSNCP-Radar-Sensing.git}{DI-GSNCP-Radar-Sensing}.}

\section{Measurement process}
\label{sec:measurement_process}

Let $\Tilde{\Phi}=\{\bs{C}_1,\dots,\bs{C}_L\}$ denote the \gls{rfs} of object states where $\bs{C}_l = (\bs{c}_l, \bs{E}_l)$ for object center positions $\bs{c}_l\in\mathcal{D}\subset \R^d$ and extents $\bs{E}_l\in\mathcal{L}_+^d$ where $\mathcal{L}_+^d$ denotes the space of $d$-dimensional lower triangular matrices with positive diagonal elements, adopting the random matrix object extent model \cite{Koch2008RandomMatrices}. The lower triangular matrices are parameterized by six parameters $\bs{e}_l=(e_{l, 1}, \dots, e_{l,6})$ and we can interpret $\bs{E}_l$ as defining an ellipse as illustrated in \cref{fig:minimal_example}. Note that the covariance matrix for the $l$-th target is then $\bs{\Sigma}_l^E \triangleq \bs{E}_l\bs{E}_l^T \in \mathcal{M}_+^d$ where $\mathcal{M}_+^d$ denotes the space of positive definite matrices.

Multiple sensors estimate the position of the objects in multiple time steps. Each set of measurements is associated with a sensor state, denoted $\bs{s}_k$ for $k=1,\dots,K$, as illustrated in \cref{fig:minimal_example}. The state encapsulates all relevant information regarding the measurement scenario.
In this paper, we consider monostatic sensing with full-duplex transceivers operating in orthogonal frequency division multiplexing with uniform planar arrays \cite{AbuShaban2018Bounds,Hyowon2022Bounce}, but the details are out of scope.


In line with standard choices for multiple extended object scenarios, as in \cite{Granstrom2018:Extended,Meyer2021:Tracking,Gilholm2005:Extended}, we now specify the measurement process. We begin by assuming that the measurement model for the $k$-th sensor state is
\begin{equation}\label{eq:measurement_equation}
    \bs{z}_{k}(\bs{C}) = \bs{v} + \bs{\varepsilon}_{k}(\bs{v}), \quad \bs{v} = \bs{c} + \bs{\varepsilon}^E
\end{equation}
where $\bs{v}$ is the effective position where the radio frequency signal impinges on the target illustrated by red crosses in \cref{fig:minimal_example}, $\bs{\varepsilon}^E \sim \mathcal{N}(\bs{0}, \bs{\Sigma}^E)$, and $\bs{\varepsilon}_k(\bs{v}) \sim \mathcal{N}(\bs{0}, \bs{\Sigma}_k(\bs{v}))$. Here $\bs{\Sigma}_k : \R^d\to \mathcal{M}_+^d$ is the sensor noise covariance mapping derived through Fisher information analysis \cite{AbuShaban2018Bounds}. Note that only an estimate of the noise covariance mapping is available due to small scale fading effects which refers to the multiplicative noise due to uncertainty in the complex channel gain.

The measurements $\Psi = \cup_{k=1}^K \Psi_k$ consists of a \gls{rfs} for each sensor state $\Psi_k = \{\bs{z}_{k, 1}, \dots, \bs{z}_{k, M_k}\}$. The number of measurements per sensor state depends on the number of targets, the extent of the targets, the detection probability and resolution for the sensor at the target positions, and the clutter intensity. Specifically, we assume that, on average, $\text{max}(1, r_k(\bs{c}_l) |\bs{E}_l|)$ positions on the $l$-th object are resolvable, as illustrated with red crosses in \cref{fig:minimal_example}, where $r_k:\R^d\to\R_+$ is the resolution function for the $k$-th sensor state. This mapping depends on the bandwidth and the size of the array aperture, and will be greatest close to the sensor. Additionally, we assume known detection probability mappings $\rho_k : \R^d \to [0,1]$ which gives the probability of detecting a resolvable point. The detection probability can be derived by solving a statistical hypothesis test problem: a point is detected with high probability if the signal power from the given point is much greater than the noise power \cite{Wymeersch2020:Detection}. Then, we define the Poisson measurement rate for the $l$-th target and the $k$-th sensor state as $\Tilde{\rho}_{k}(\bs{C}_l) = \rho_k(\bs{c}_l) \text{max}(1, r_k(\bs{c}_l) |\bs{E}_l|)$. 
Finally, the clutter measurements are assumed to follow a Poisson point process with intensity $\lambda^c$.
\section{Doubly inhomogeneous-generalized shot noise Cox process}
\label{sec:model}

Assuming $\bs{\Sigma}_k(\bs{c}) \approx \bs{\Sigma}_k(\bs{v})$, we may express \cref{eq:measurement_equation} as a Gaussian stochastic process $\bs{z}_{k}(\bs{C}) \sim \mathcal{N}(\bs{c}, \Tilde{\bs{\Sigma}}_{k}(\bs{C}))$ where $\Tilde{\bs{\Sigma}}_{k}(\bs{C}) \triangleq \bs{\Sigma}^E + \bs{\Sigma}_k(\bs{c})$. 
This motivates to assume $\Psi_k|\Tilde{\Phi}$ is Poisson with intensity
\begin{equation}\label{eq:DI-GSNCP}
    Z^{(k)}_{\Tilde{\Phi},\lambda^c}(\bs{\xi}) = \lambda^c + \sum_{l=1}^{L} \eta_k(\bs{\xi} | \bs{C}_{l})
\end{equation}
where $\eta_k(\bs{\xi} | \bs{C}_{l}) = \Tilde{\rho}_k(\bs{C}_l) \mathcal{K}_{\Tilde{\bs{\Sigma}}_k(\bs{C}_l)}(\bs{p}_{k, m_k} - \bs{c}_l)$ and $\mathcal{K}_{\bs{\Sigma}}$ is the Gaussian kernel function with bandwidth $\bs{\Sigma}$, i.e.,
\begin{equation*}
    \mathcal{K}_{\bs{\Sigma}}(\bs{\xi}-\bs{c}) = \frac{\exp\big(-\frac{1}{2}(\bs{\xi}-\bs{c})^T\bs{\Sigma}^{-1}(\bs{\xi}-\bs{c})\big)}{\sqrt{(2\pi)^d|\bs{\Sigma}|}}.
\end{equation*}
We say that $\Psi_k$ is a doubly inhomogeneous-generalized shot noise Cox process (DI-GSNCP) where it is doubly inhomogeneous since inhomogeneity is introduced in both the cluster size and cluster spread parameters as well as the kernel function and it is generalized shot noise since the driving process includes a parameter that changes the cluster size and spread \cite{Moeller2005:GeneralisedShotNoise,Dvorak2022:binspp}.
We assume further that $\Phi=\{\bs{c}_1,\dots,\bs{c}_L\}$ is a hard core process with intensity $\lambda>0$ and known hard core radius $R>0$. Finally, we assume that $\bs{\Sigma}^E_l$ are independent and identically distributed.

\subsection{Posterior distribution}


Consider now the model for the \gls{rfs} of the observed measurements conditioned on the target states, i.e., $\Psi_k|\Tilde{\Phi}$, with intensity given in \cref{eq:DI-GSNCP}. The model parameters are $\theta=(\Tilde{\Phi}, \lambda, \lambda^c)$ and, up to a constant, the posterior is $\Pi(\theta | \Psi) = \prod_{k=1}^K \Pi_{\text{likelihood}}(\Psi_k|\Tilde{\Phi}, \lambda^c) \Pi_{\text{prior}}(\Tilde{\Phi}, \lambda, \lambda^c)$.
Using the Poisson assumption, the observed likelihood is
\begin{equation*}
    \begin{split}
        \Pi_{\text{likelihood}}(\Psi_k|\Tilde{\Phi}, \lambda^c) &= \exp((1-\lambda^c)|\mathcal{D}| - \sum_{l=1}^{L} \Tilde{\rho}_k(\bs{c}_l))\\
        &\qquad \times \prod_{m_k=1}^{M_k} Z^{(k)}_{\Tilde{\Phi}, \lambda^c}(\bs{p}_{k, m_k}).
    \end{split}
\end{equation*}
The prior is proportional to
\begin{equation*}
    \begin{split}
        \Pi_{\text{prior}}(\Tilde{\Phi}, \lambda, \lambda^c) &= \lambda^{L} \mathbbm{1}_{\R_+}(\lambda) \mathbbm{1}_{\R_+}(\lambda^c) \prod_{l=1}^L \Pi_E(\bs{\Sigma}^E_l)\\
        &\qquad \times \prod_{\substack{j=1\\j\neq l}}^L \mathbbm{1}_{\R_+}(\Vert \bs{c}_l - \bs{c}_j \Vert - R)
    \end{split}
\end{equation*}
where $\mathbbm{1}_A(x)$ is the indicator function which equals $1$ if $x\in A$ and $0$ otherwise, $\Pi_E$ is the extent prior density, and we have assumed uniform priors on the cluster center and clutter intensities.

\section{Jump Markov chain Monte Carlo}
\label{sec:markov}

The posterior is only known up to proportionality which motivates working with Metropolis-Hastings \gls{mcmc}.
Using \gls{mcmc}, a Markov chain $\theta_0, \theta_1,\dots$ is constructed as follows:
given the previous configuration in the Markov chain, i.e$.$, $\theta_{i-1}$, a possible new configuration of model parameters is sampled as $\theta^*\sim\mathcal{Q}_{(\theta_{i-1})}(\cdot)$ from a predefined transition density. We define the acceptance probability function $\alpha(\theta_{i-1}, \theta^*) = \text{min}\Big\{1, \frac{\Pi(\theta^*|\Psi)}{\Pi(\theta_{i-1}|\Psi)}\Big\}$ and sample $U\sim\text{Unif}(0,1)$. If $U < \alpha(\theta_{i-1}, \theta^*)$ set $\theta_i = \theta^*$, otherwise let $\theta_i = \theta_{i-1}$.

In this work, the parameters in $\theta$ are updated sequentially following the ideas of \cite{Kopecky2016Bayesian}. Firstly, the driving process $\Tilde{\Phi}$ is updated followed by an update of $\lambda$ and $\lambda^c$. We define the transition density for updating the driving process as a birth-death-move proposal: with probability $p_m$ a move is taken, with probability $(1-p_m)p_b(\theta_{i-1})$ a birth is proposed, and with probability $(1-p_m)(1-p_b(\theta_{i-1}))$ a death is proposed. Specifics of the proposed \gls{mcmc} sampling technique and subsequent target state estimation is given in the following:

\emph{\textbf{Birth update}}:
we design the birth rate as \cite[p.~152]{Lieshout2000}
\begin{equation*}
    b(\theta_{i-1}, \bs{\Xi}) = \lambda_{i-1} \bigg(1 + \sum_{k=1}^K\sum_{m_k=1}^{M_k} \frac{\eta_{k}(\bs{p}_{k,m_k} | \bs{\Xi})}{\lambda_{i-1}^c}\bigg),
\end{equation*}
for $\bs{\Xi} \in \mathcal{D} \times \mathcal{L}_+^d$ where the birth density is defined after normalization by the total birth rate $B(\theta_{i-1}) = \int b(\theta_{i-1}, \bs{\Xi}) d\bs{\Xi}$. For computational purposes, we only compute the birth rate on a discretization of $\mathcal{D} \times \mathcal{L}_+^d$ and define a non-uniform discretization of $\mathcal{D}$ which focuses computations on areas of the domain where the density of measurements are high. Specifically, the discretization of $\mathcal{D}$ is an independent and homogeneous thinning of $\Psi$ with retention rate $\text{min}(1,\frac{N_B}{|\Psi|})$, where $N_B\in\N$ is the expected cardinality of the discretization of $\mathcal{D}$, while the discretization of $\mathcal{L}_+^d$ is constructed by sampling $N_E$ extent parameters from $\Pi_E$. Then we define the birth rate on the union of balls with radius $r_B$ centered at points in the constructed discretization.

\emph{\textbf{Death update}}: we define the death rate to fulfil the detailed balance condition $d(\theta_{i-1} \cup \bs{\Xi}, \bs{\Xi}) = \frac{b(\theta_{i-1}, \bs{\Xi}) \Pi(\theta_{i-1} | \Psi)}{\Pi(\theta_{i-1} \cup \bs{\Xi} | \Psi)}$, abusing the notation $\theta_{i-1} \cup \bs{\Xi} \equiv (\Tilde{\Phi}_{i-1} \cup \bs{\Xi}, \lambda_{i-1}, \lambda_{i-1}^c)$. The total death rate $D(\theta_{i-1}) = \sum_{\bs{C}_l \in \Tilde{\Phi}_{i-1}} d(\theta_{i-1}, \bs{C}_l)$ is the normalization factor for the death transition density. The birth probability is chosen as $p_b(\theta_{i-1}) = \frac{B(\theta_{i-1})}{D(\theta_{i-1})+B(\theta_{i-1})}$ \cite[p.~86]{Lieshout2000}.

\emph{\textbf{Move update}}:
to alleviate the difficulties of Markov chain transitions in high-dimensional spaces, we keep move steps of the target position and the target extent separate. When a move update is to be made, with probability $1-p_{em}$ (resp. $p_{em}$) a move is made for the position (resp. extent). We define the move transition density for moving the spatial position of the $l$-th cluster center as a Gaussian density with mean $\bs{c}_l$ and covariance $\vartheta \Big(\sum_{k=1}^K \Tilde{\bs{\Sigma}}_k^{-1}(\bs{C}_l)\Big)^{-1}$, $\vartheta >0$. In case of an extent move, the transition for the extent parameters follows a Gaussian density with mean $\bs{e}_l$ and covariance $\sigma^2_{E, \text{move}} \text{diag}(\bs{e}_l^2)$, $\sigma_{E, \text{move}}>0$. To make frequently accepted large steps, $\vartheta$ and $\sigma_{E, \text{move}}$ are adaptively adjusted: if a move (resp. extent move) is rarely accepted then decrease $\vartheta$ (resp. $\sigma_{E, \text{move}}$), and vice versa.

\emph{\textbf{Clutter/non-clutter labelling}}:
in the model intensities update, we rely on labelling the measurements as clutter or non-clutter. Hence, in each iteration of the Markov chain, a clutter association hypothesis is sampled as $\{a_{k,1},\dots,a_{k,M_k}\}$ where $a_{k,m}=0$ if clutter and $1$ otherwise. This sampling can be done according to the conditional density of clutter/non-clutter labelling as
\begin{equation*}
    \begin{cases}
        \frac{\lambda_{i-1}^c}{Z^{(k)}_{\Tilde{\Phi}_{i},\lambda_{i-1}^c}(\bs{p}_{k, m_k})}, ~ &\text{if } a_{k,m_k} = 0,\\
        \displaystyle\sum_{l=1}^L \frac{\eta_k(\bs{p}_{k, m_k} | \bs{C}_l)}{ Z^{(k)}_{\Tilde{\Phi}_{i},\lambda_{i-1}^c}(\bs{p}_{k, m_k})}, ~ &\text{if } a_{k,m_k} = 1.
    \end{cases}
\end{equation*}

\emph{\textbf{Model intensities update}}:
for simplicity, in the $i$-th configuration of the Markov chain, we define the update of the cluster center and clutter intensities as
\begin{equation*}
    \lambda_i = \frac{(|\Psi|-M^c_i)~|\Phi_i|}{|\mathcal{D}|~\sum_{C_l\in\Tilde{\Phi}}\sum_{k=1}^{K} \Tilde{\rho}_k(\bs{C}_l)} \quad \text{and} \quad
    \lambda^c_i = \frac{M^c_i}{K|\mathcal{D}|},
\end{equation*}
respectively, where $M^c_i$ are the number of measurements labelled as clutter. These choices have shown to perform well in practice \cite{Kopecky2016Bayesian}.

\emph{\textbf{Target state estimation}}:
following the conclusion of the \gls{mcmc} algorithm, the estimates of the target states are
\begin{equation*}
    \Tilde{\Phi}^\star = \{\bs{C}_1^\star,\dots,\bs{C}_{L^\star}^\star\}, \quad \bs{C}_l^\star = \frac{1}{N} \sum_{i=N_0}^{N_0+N} (\bs{C}_l)_i.
\end{equation*}
where $N_0$ are the number of burn-in iterations and $N_0+N$ is the total number of \gls{mcmc} iterations. The number of burn-in iterations is adaptively chosen by an early-stopping strategy: when the posterior has not increased for a number of iterations specified by a patience parameter, end the burn-in. We note that after the burn-in, the move probability $p_m$ is set to one such that the cardinality remains constant.

\vspace{-0.2cm}
\section{Numerical experiments}
\label{sec:numerical}

We consider two baseline methods: (i) \textit{Oracle} where we assume perfect knowledge of the data association and the sensor noise covariance --- for target position estimation, we further assume that the true object extents are known, while for the extent process, we assume that the true object locations are known, allowing tractable maximum likelihood solutions; (ii) \textit{DBSCAN} where we use the DBSCAN clustering algorithm as a low complexity state-of-the-art method. The hyperparameters are optimized by a grid search, maximizing the posterior.

To compare sensing accuracy, we use the optimal sub-pattern assignment (OSPA) metric \cite{Schuhmacher2008:OSPA} with pair-wise metric defined by the Gaussian Wasserstein distance as recommended in \cite{Yang2016:Metrics}. The order and cut-off of the OSPA metric is set to $2$ and $10$, respectively.

We define a scenario with two sensors scanning over $6$ time epochs. The spatial domain is $\mathcal{D}=[0, 50]\times[0, 20]\times[0, 10]~m^3$ and we assume extent prior as $e_{l,1},e_{l,2},e_{l,3} \sim \text{Unif}(1, 1.5)$ and $e_{l,4},e_{l,5},e_{l,6} \sim \text{Unif}(-0.5, 0.5)$. Moreover, we let $R=8~m$, $\lambda=\frac{20}{|\mathcal{D}|}$, and $\lambda^c=\frac{35}{|\mathcal{D}|}$.

The move and extent move probabilities for the \gls{mcmc} algorithm are fixed as $p_m=0.8$ and $p_{em}=0.7$. The birth transition hyperparameters are fixed as $N_B=100$, $N_E=20$, and $r_B=0.2$, the patience parameter $200$, and $N=100$. Regarding initialization, at the initial time epoch, the Markov chain is initialized from a birth step, and at the subsequent time epochs, the Markov chain is initialized by the estimate from the previous time epoch.


\begin{figure}[t!]
    \centering
    \begin{minipage}{0.47\textwidth}
        \centering
        \resizebox{0.8\textwidth}{!}{\input{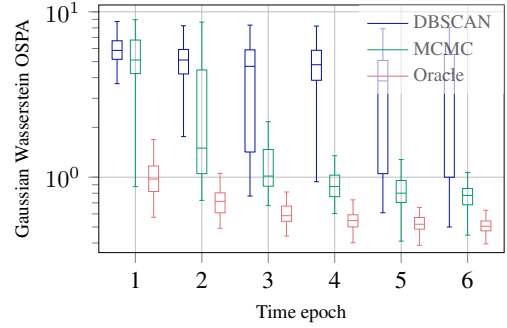}}
    \end{minipage}\\
    \vspace{-0.20cm}
    \caption{Distribution of the sensing accuracy.}
    \vspace{-0.30cm}
    \label{fig:ospa_boxplot}
\end{figure}

In \cref{fig:ospa_boxplot}, the empirical distributions, evaluated with $100$ Monte Carlo simulations, of the sensing accuracy, in terms of the Gaussian Wasserstein OSPA distance for the \gls{mcmc} algorithm and the two baseline methods, are summarized by boxplots as the number of scans increase. We see that the \gls{mcmc} algorithm significantly outperforms the DBSCAN baseline after just two time epochs. Moreover, the \gls{mcmc} algorithm is close to reaching the lower bound provided by the oracle after four time epochs after which the accuracy saturates. In this experiment, the number of \gls{mcmc} iterations, averaged across the Monte Carlo simulations, is below $1000$ in all time epochs, showcasing the low complexity of the algorithm.

\section{Conclusion}
\label{sec:conclusion}

A methodology for target state estimation in multi-sensor multi-scan multiple extended target sensing scenarios is developed. The method is based on parametrizing the target states through a doubly inhomogeneous-generalized shot noise Cox process taking spatial properties of multiple sensors into account and using a jump Markov chain Monte Carlo algorithm to estimate the parameters. The method scales only linearly in the number of measurements, effectively estimating the target states without requiring data association. Numerical experiments demonstrate the benefits over spatial proximity based clustering in high clutter scenarios with closely spaced targets.

In future work, we aim to generalize the method to non-linear measurement models in scenarios with moving targets.






\vfill\pagebreak

\bibliographystyle{IEEEbib}
\small
\bibliography{bib}

\end{document}